\documentclass[journal,onecolumn]{IEEEtran}
\IEEEoverridecommandlockouts
\usepackage{cite}
\usepackage{amsmath,amssymb,amsfonts}
\usepackage{algorithmic}
\usepackage{graphicx}
\usepackage{textcomp}
\usepackage{xcolor}
\usepackage{verbatim}
\usepackage{listings}
\usepackage{adjustbox}
\usepackage{booktabs}
\usepackage[utf8]{inputenc}
\usepackage{newunicodechar}
\newunicodechar{⁴}{\textsuperscript{4}}

\begin{document}
%


\title{Can Open-Source LLM Agents Replace Static Application Security Testing Tools? An Empirical Assessment}

%

\author{

Derek Yohn¹, Luke Flancher², Mirajul Islam³, and Khaled Slhoub⁴ \\
College of Engineering and Science, Florida Institute of Technology,
Melbourne, Florida, USA \\
¹dyohn2025@my.fit.edu, ²lflancher2026@my.fit.edu, 
³islamm2024@my.fit.edu, ⁴kslhoub@fit.edu

}

\maketitle

\begin{abstract}
This paper explores the value of agentic AI tools for cybersecurity purposes. We evaluate the efficacy of a general-purpose GenAI Large Language Model- (GenAI-) based agent when powered by three different Ollama-hosted general-purpose open source models. We assess each agent's performance using precision, recall, false positive count, and a calculated composite score based upon the interplay of the captured metrics, against the baseline performance of an existing, vetted Static Application Security Testing (SAST) tool, Bandit. Our findings refute the notion that a modern open-source GenAI LLM-based agent is currently suitable for the specialized task of SAST scanning under realistic conditions.
\end{abstract}


\section{Introduction} \label{sec:introduction}


Since 2023, LLMs and the field of artificial intelligence has experienced explosive progress, with the technology being almost unrecognizable today compared to the very same technology three years ago \cite{durrani_decade_2024}. Each iteration has introduced new capabilities and productivity promises across numerous industries and job fields.  What began as simple chatbots \cite{sachdeva_taking_2024} have become generative (GenAI) systems, capable of creating or summarizing text, code, images, audio, and even videos based on user-provided prompts. These new systems are a far cry from the single-purpose expert systems that experienced their heyday in the 1980s \cite{durrani_decade_2024}. Rather, they are intended to be general-purpose systems, imbued with the collective knowledge of the internet, digital libraries, and publicly-available software repositories. The introduction of attention mechanisms \cite{wu_attention_2025}, coupled with massively-scaled compute backends \cite{krizhevsky_imagenet_2017} processing potentially millions of tokens, even helped give rise to the not-so-subtle hype that artificial general intelligence (AGI) \cite{maruyama_conditions_2020}, intelligence that rivals or surpasses what a skilled human could produce, could be within reach. The veracity of such claims notwithstanding, GenAI does offer the potential to radically alter how humans work. 

From summarizing emails to generating art or humor, no professional practice has been as dramatically impacted as that of software engineering \cite{aguilar-lopez_impact_2025}. It is now not uncommon for software to be written by GenAI systems, guided by a skilled practitioner. Thus far, the pure generation of such code has been the hot topic in both academia and industry. While code templating and autocomplete have been around for many years, modern GenAI systems such as Claude Code \cite{chatlatanagulchai_use_2026, santos_decoding_2025} have brought code generation out of the integrated development environment (IDE) and into a collocated deployment. In the case of Claude Code, this is a local terminal session. Other tools, such as Ollama \cite{srinivas_building_2025, noauthor_ollamas_nodate}, allow the creation of what are known as agents. These agents are independent processes or systems, capable of autonomously providing some type of GenAI service on-demand. Such a model is known as agentic AI, and has commonly been used for the creation of code. We, however, are interested not in the creation of code, but in its validation, specifically its cybersecurity fitness.

Numerous tools for evaluating the cybersecurity fitness of code exist \cite{kuszczynski_comparative_2023}. Typically, such tools are either manually run by the end user, or included in some type of build/deploy pipeline \cite{bennett_developers_2024}. Cybersecurity scanning tools are divided into two types: static application security testing (SAST) and dynamic application security testing (DAST). SAST tools analyze code that is not executing, seeking to identify problems such as potential SQL injections, poor secret management, or insecure coding practices \cite{li_comparison_2023}. DAST tools examine the behavior of software during runtime, such as improperly configured HTTP headers, business logic errors, and server-side request forgeries \cite{stanciu_security_2025}.

The purpose of this research paper is to explore the value of agentic AI tools for SAST cybersecurity purposes. More specifically: can LLM-based cybersecurity scanning agents rival or surpass traditional tooling? An agentic approach to SAST testing complements recent advances in AI-assisted code generation because generated source code can include hallucinations and other security vulnerabilities. Further, many individual developers or small/medium teams may not have achieved sufficient process maturity, meaning that they do not typically have SAST tooling fully integrated into their workflow. Were such an agent proven effective, those types of development teams could leverage the power and simplicity of an agentic security scanner that watches their codebase as it evolves, providing feedback quickly on-demand or perhaps even automatically. In essence, such an agent would be capable of Agentic pair programming, improving the overall quality of the already-written or generated code with human expert guidance. The main contribution of this paper is an empirical evaluation of a simple SAST agent, providing evidence that a modern general-purpose GenAI LLM-based agent is not yet suitable for the specialized task of SAST scanning under realistic conditions. Performance efficacy of our created agent is empirically measured. Using the output of a known and vetted open-source tool as the baseline of comparison, we capture our agent's precision, recall, false positive count, and calculate a composite score based upon the interplay of the captured metrics.

In Section~\ref{sec:background} we discuss AI/ML concepts, trends, and developments that have made agentic AI systems possible. In Section~\ref{sec:relatedWork}, we examine relevant similar studies and how we differentiate our work. Section~\ref{sec:methodology} presents our experimental goals and design. Experimental outcomes are reflected upon in Section~\ref{sec:results}, while a detailed discussion of the findings is provided in Section~\ref{sec:discussion}. Threats to experimental validity are addressed in Section~\ref{sec:threatsToValidity}. We offer a final summary and closing thoughts in Section~\ref{sec:conclusion}.

\section{Background} \label{sec:background}

The field of artificial intelligence is undergoing a transformative shift from task-specific assistive tools toward agentic AI, a paradigm defined by autonomous systems capable of pursuing complex goals with minimal human intervention \cite{acharya_agentic_2025}. Unlike traditional AI, which operates under strict supervision and predefined boundaries, agentic AI demonstrates a "qualitative leap" in autonomy, enabling systems to operate dynamically in uncontrolled real-world environments \cite{acharya_agentic_2025}. These systems are characterized by their goal-directed behavior, dynamic adaptation, and self-improvement capabilities \cite{pati_agentic_2025}. This evolution has been fueled by the maturation of Generative AI, which has transitioned from early probabilistic models to high-fidelity architectures like Transformers and Diffusion Models, revolutionizing content creation across text, images, and structured data \cite{trigka_evolution_2025}.

The integration of these advanced models into cybersecurity presents a dual-use landscape of risks and opportunities: while Large Language Models (LLMs) empower defenders to perform real-time threat detection and zero-day vulnerability scanning, they are simultaneously exploited by adversaries to generate automated malware and obfuscate malicious code \cite{ahi_large_2025}. AI is moving the industry from reactive to proactive strategies, utilizing models like Random Forest to predict cyber-attacks with high accuracy by sensing and reasoning through unrecognized threats \cite{ankalaki_cyber_2025}. 

The significance of Machine Learning (ML) and Deep Learning (DL) in this domain is critical, as these techniques provide the necessary frameworks for foreseeing attacks and revealing root causes through the analysis of malware traffic and system logs \cite{mijwil_significance_2023}. Unknown patterns of data can be analyzed via unsupervised, supervised, reinforced, and algorithmically-based ML to improve cybersecurity \cite{alshuaibi_machine_2025}. Furthermore, the emergence of conversational agents like ChatGPT has introduced new mechanisms to both assist in creating safe digital environments and challenge current security practices through sophisticated fraudulent operations \cite{mijwil_towards_2023}. 

In software engineering, researchers envision a future defined by a symbiotic partnership between human developers and AI to boost productivity across the entire development lifecycle \cite{terragni_future_2025}. However, the widespread adoption of AI-generated programming code has introduced significant security implications: because AI models are often trained on large code repositories containing insecure components, they frequently suggest code with Common Weaknesses Enumeration (CWE) flaws \cite{dimitrov_cybersecurity_2025}. Integrating Generative AI into Source Code Management (SCM) systems further exacerbates these risks, leading to threats such as involuntary vulnerability introduction, data poisoning, and supply chain risks \cite{nethala_cyber_2025}. 

To address these challenges, specialized prompting techniques for secure code generation, such as Recursive Criticism and Improvement (RCI), have been developed to guide LLMs toward producing more secure implementations by leveraging the models' self-critiquing capabilities \cite{tony_prompting_2025}. Additionally, hybrid frameworks like the ANN-ISM paradigm provide a road map for organizations to improve their security maturity by combining the pattern-recognition power of neural networks with hierarchical risk mitigation strategies \cite{khan_ai-driven_2025}. The transformation of vulnerability management is also underway through AI-driven models that move beyond syntactic pattern matching to semantic reasoning, utilizing risk-aware prioritization to combat "alert fatigue" in large-scale software estates \cite{siewruk_transformation_2026}. These complex, decentralized operations are often supported by Multi-Agent Systems (MAS), where autonomous entities collaborate and coordinate to solve tasks that are beyond the capacity of individual agents \cite{dorri_multi-agent_2018}.

It is increasingly popular to use lightweight vulnerability techniques like static analysis for security verification, and there is an expectation that software engineering will shift its focus from design and implementation to evolution, maintenance, and security-hardening \cite{bohme_software_2025}, a role that agentic AI may be well-suited to. One defining characteristic of agentic AI is its ability to make decisions based on its observed environment in order to execute multi-step tasks to achieve its defined goal. Such an approach is powerful, but can expose many attack vectors such as prompt injection; collocation of Agents via small models such as Ollama can mitigate much of that risk while still providing the same benefits \cite{obeng_security_2025}. LLM-based agents offer a new approach to building hybrid intelligence workflows where they act as a type of brain for traditional tools, and they show great promise for the creation of high-quality synthetic data for testing and training purposes \cite{xu_large_2025}. More importantly, AI tools can significantly increase efficiency by automation of manual coding and documentation \cite{russo_navigating_2024}, with the key to their adoption being perceived usefulness, ease of use, and compatibility with current workflows.

\section{Related Work} \label{sec:relatedWork}



Agentic cybersecurity is an emerging field of interest. Consequently, direct research on the topic is lacking. However, other researchers have examined the cybersecurity aspect of GenAI (LLM) coding from a variety of angles. While not quite strictly empirical works such as ours, these related works do offer relevant guidance.

A popular approach to cybersecurity is to directly apply traditional AI/ML implementations for the purpose of pattern-recognition and classification. These can vary from direct application of classic machine or deep learning to hybrid systems. The overarching theme is the use of the known system for a new goal.

Wang et al. contribute a quintessential approach, applying LLMs for web testing \cite{wang_enhanced_2026}. They find that off-the-shelf models simply lack a deep understanding of complex web page structures, let alone private business logic. Crucially, they replicate the well-known hallucination vulnerability of LLMs which are the source of developer mistrust. Mitigations are proposed, but this is an inherent problem.

Khan et al. \cite{khan_ai-driven_2025} provide an interesting hybrid approach, leveraging the power of Artificial Neural Networks (ANN) hybridized with the hierarchical structuring of Interpretive Structural Modeling (ISM) to develop a framework designed to improve secure software coding. Their model could classify risks into 15 different specific categories, with "Insecure Coding Practices" and "Vulnerable Dependencies" identified as high impact. Unsurprisingly, there were the usual challenges: training and use of their framework is computationally expensive, suffers from the "black box" problem of unexplainability, and struggles to adapt to novel, unknown attack techniques. Their framework provides a maturity roadmap for organizations and integration points for other AI paradigms such as reinforcement learning and genetic algorithms. 

Millar et al \cite{millar_optimising_2022} also contribute a hybrid system, leveraging a deep learning architecture to optimize vulnerability triage in DAST settings. They combine multiple Convolutional Neural Networks (CNNs) and Natural Language Processing (NLP) to classify scans results as either verified vulnerabilities or false positives. The false positives are problematic, but were reduced by 20\% by their work. However, most companies do not deploy defensive AI/ML in any form. This work touches on a topic (DAST) that is tangential to our work, and does so in a classical way rather than our empirical approach. Xu et al. \cite{xu_seek_2026} also propose a hybrid system, in the form of a reasoning pipeline for Digital Risk Protection (DRP), which blends multi-modal LLMs with human-curated references. They again find hallucinations to be a source of troubles. Most relevant to our approach, they note that human references act as a decision-time override for AI suggestions/hypotheses. 

Sheng et al. \cite{sheng_llms_2025} take a classical LLM tack towards vulnerability detection, but ultimately find that the community has shifted from traditional CNN/RNN models to LLMs. Further, multi-agent approaches are widely used to decompose complex security challenges; this reflects our understanding of the current trends in GenAI usage. They identified major research gaps such as a narrow research scope focused on isolated code snippets rather than full repositories and the widespread problem of data leakage inflating performance metrics. The fine-tuning mechanisms for frontier LLMs (e.g., GPT-4 and Claude 3.5) provide Chain-of-Thought mechanisms that may be effective mitigations, though more research is required. 

Moving away from classical neural network architectures, some researchers directly examine the cybersecurity fitness of code generated by LLMs (GenAI). Klemmer et al. \cite{klemmer_using_2024} tackle code generation right at that source, qualitatively studying how software developers use AI assistants like ChatGPT and GitHub Copilot as their GenAI tools of choice. Such tools have experienced rapid adoption for tasks like code generation ethical security investigations. Developers often view AI as a novel source of advice as compared to Stack Overflow or search engines; this strikes us as strange, given that the LLMs are trained on exactly such sources. Yet again, hallucinations are a scourge, potentially leading to vulnerabilities.

Perhaps it is the prompting of the LLM that is a source of misuse, rather than the LLM's generation itself? Tony et al. \cite{tony_prompting_2025} examine just this topic, evaluating 15 prompting techniques to determine their impact on the security of GenAI code. They ultimately find that there are some optimization strategies that can improve the generated code from user prompts. However, this is tedious and requires extensive security knowledge. 

It turns out that it is not simply prompting that introduces vulnerabilities, it is the code snippets generated by models like Copilot and Meta's CodeLlama, as discovered by Nethala et al \cite{nethala_cyber_2025}. Models can actually hallucinate the existence of entire packages, as well as introduce code vulnerabilities of their own. Human oversight is the only reasonable oversight mechanism at this time.

Agent-based GenAI systems have very little research attached with respect to cybersecurity concerns. It was established by Wadinambiarachchi et al \cite{wadinambiarachchi_imagining_2025} that agentic AI has the capacity to transform creative design workflows by acting as strategic, goal-oriented partners rather than mere tools. The danger, of course, is that designers may come to rely too heavily on AI suggestions, particularly early in the design process. Petrovic et al. \cite{petrovic_agent-based_2025} studied an agentic approach to Internet of Things (IoT), from both offensive and defensive perspectives. They noted the complexity of the topic, and ultimately relied upon design-time prompting and runtime DevSecOps and Chain-of-Thought reasoning to make security assistants valuable. Overall, research in this facet is extremely limited. However, we can say that agentic AI can serve a helpful purpose in an assistant role, and that applications to cybersecurity are promising from multiple perspectives.

More directly aligned with our investigation is the creation of custom tools for cybersecurity purposes. There were attempts by Ming'ate and K to derive security scanners for Cross-Site Vulnerabilities \cite{mingate_ai-driven_2025}, resulting in a tool they named "VulScanWare." This was an AI-augmented multi-stage scanner, combining web crawling, vulnerability identification, and AI-driven repair. Pendyala and Jeswani \cite{pendyala_zt-icas_2026} contribute the first agentic method we discovered (ZT-ICAS), a zero-trust integrity-constrained framework for the purpose of security scanning. Such a system is powerful, but requires robust architectures to scale for real-world use, not to mention that it still suffers from prompt injection and data poisoning, particularly when not supervised by a human expert.

Yang et al. \cite{yang_knighter_2025} provide the tooling most similar to our goals. They introduce KNighter, an LLM-base framework for synthesizing static analysis checkers for OS kernels. Their focus is on bug patterns in massive codebases like the Linux kernel, while we aim to sit at the level of the individual developer, watching them work and acting as a potential gatekeeper. Notably, they identify a multi-stage synthesis pipeline as a potential solution to the problem of scale, something we may need consider.

 Finally, there are two empirical studies of note. First, Kapetanidou et al. \cite{kapetanidou_evaluation_2025} explicitly examine available security scanning tools, though they focus on the Kubernetes and edge-cloud ecosystem. Security misconfigurations are identified as high risk, and important industry tools such as Trivy, Kubescape, and Checkov are evaluated. While their focus on specific security tooling is similar to our aims, the narrow focus on Kubernetes clashes with our view of developer-level testing. Second, and much more interestingly, Al-Shammare et al. \cite{al-shammare_evaluating_2025} provide the study most similar to our efforts. They evaluate the effectiveness of emerging SAST tools, such as Fortify-SCA, Sparrow SAST, and PVS-Studio, within the Software Development Life Cycle (SDLC). They recognize that SAST tools become increasingly vital as codebases grow in complexity, as they can detect vulnerabilities early and help reduce project risk. Further, they note the gap in existing literature regarding the comparison of tool outputs for defect detection and limited research on tool usability, recommending automated scans to improve efficiency. This is closely related to the potential impact of our work.

While non-trivial research has been conducted with respect to the intersection of GenAI, agents, and cybersecurity, there exists no study with the same goals and methodologies as ours. We wish to examine the efficacy of using agentic AI as a developer assistant, assuming the responsibilities normally assumed by separate SAST tooling. Ours is a novel application of agentic AI, but one which requires empirical comparison and analysis.

\section{Methodology} \label{sec:methodology}

To study how AI tools compare to traditional rule-based tools in cybersecurity, we pitted the two against each other. Given the dearth of research in this burgeoning field, we set out to establish a minimal baseline of comparison. It is simply impossible to exhaustively test the numerous SAST tools available. They are far too numerous, and many are not open-source and require enterprise licensing. Our goal is to evaluate the performance of a general-purpose local LLM agent in identifying cybersecurity vulnerabilities within codebases, thus we only require a reliable comparative results baseline from a known and vetted tool. To form our qualitative basis of comparison, we selected an existing SAST cybersecurity tool named Bandit \cite{kuszczynski_comparative_2023, noauthor_pycqabandit_2026}. Bandit, a Python-focused tool, has a low rate of false positives and generally performs moderately well in terms of accuracy, precision, and sensitivity \cite{kuszczynski_comparative_2023}, making it a reasonable basis of comparison for our Python-coded SAST Agent. It identifies potential threats on a scale of low, medium, and high, coupled with confidence levels of low, medium, and high. Bandit draws upon MITRE's Common Vulnerabilities and Exposures (CVE) definition dictionary \cite{noauthor_mitre-cyber-security-cve-databasemitre-cve-database_2026} to assist in vulnerability detection, and provides a link to the relevant CVE for each detection. Output is available in several formats; we chose JSON as the common output format to support machine-based parsing and ease of review.

We next selected the computing environment where Ollama models would be hosted. That same environment was used for all Bandit and agent scans conducted. Bandit does not advertise any minimum system requirements, but is a small binary that parses code into efficient abstract syntax trees for analysis. Regardless, Bandit's computational needs will be irrelevant as compared to even modest token generation and consumption of even the smallest LLM. For hosting of local LLMs, Ollama's documentation \cite{noauthor_ollamas_nodate} recommends at least a modern CPU (four or more cores), 16 GB RAM, an NVIDIA GPU series with at least 8 GB VRAM, and a disk, preferably solid state (SSD), with 4 GB for Ollama itself plus another 10-50 GB for models. Our hosting/testing environment greatly exceeded those recommended specifications, using an Alienware Gaming Laptop designed for heavy computation and rendering via an Intel Core Ultra 9 Series 1 chip (16 cores at 5.1 GHz CPU Base Clock Frequency) and an NVIDIA GeForce RTX 4070 with 12 GB VRAM. A full environment specification comparison can be seen in Table~\ref{tab:techSpecs}.

\renewcommand{\arraystretch}{1.5}
\begin{table}[hbtp]
    \centering
    \caption{Technical Specification of the Ollama Hosting and Test Environment}
    \begin{tabular}{l|cr}
         & Recommended & Actual  \\
         \toprule
         CPU Cores & 4 & 16 \\
         RAM (GB) & 16 & 64 \\
         VRAM (GB) & 8 & 12 \\
         Disk Type & SSD & SSD \\
         Disk Size (GB) & 50 & 2000 \\ 
         \bottomrule
    \end{tabular}
    \vspace{0.5em}

    \label{tab:techSpecs}
\end{table}
\renewcommand{\arraystretch}{1}

With the baseline SAST tool and testing environment selected, we turned our attention to the problem of what to scan. For consistency, and to conform with Bandit's use case, it was decided that any scanning target need be an open source software package written primarily in the Python programming language. Using open source was cost effective, and the choice of Python both provides many candidate tools/packages and aligns with the coding environment used to create our agent, to be discussed momentarily. A search of GitHub eventually yielded three repositories that matched our needs. Each repository houses a tool geared towards a different use case, providing us some variety in tool purpose and quality. The first tool selected is the now-retired \textbf{Yum} package manager for Linux rpm-based systems \cite{noauthor_rpm-software-managementyum_2026}. Yum had a long service life in systems such as RedHat and Fedora, and represents enterprise-ready, open source software of the highest expected quality. The second tool selected is called \textbf{Beaverhabits} \cite{zhu_daya0576beaverhabits_2026}, a personal habit tracker originally designed for mobile devices, but which can be self-hosted in a Docker container for desktop use. This tool is "casual commercial," if you will, and cross-platform. The third and final software tool selected for scanning was a security-oriented service daemon named \textbf{Fail2ban}. This tool aims to keep a user's Linux-based system safe from penetration by proactively banning connection attempts (e.g., SSH) from IP addresses who have conducted too many failed login attempts, a feat it achieves via scanning of relevant log files and manipulation of firewall rules.

\begin{figure}
    \centering
    \includegraphics[width=0.9\linewidth]{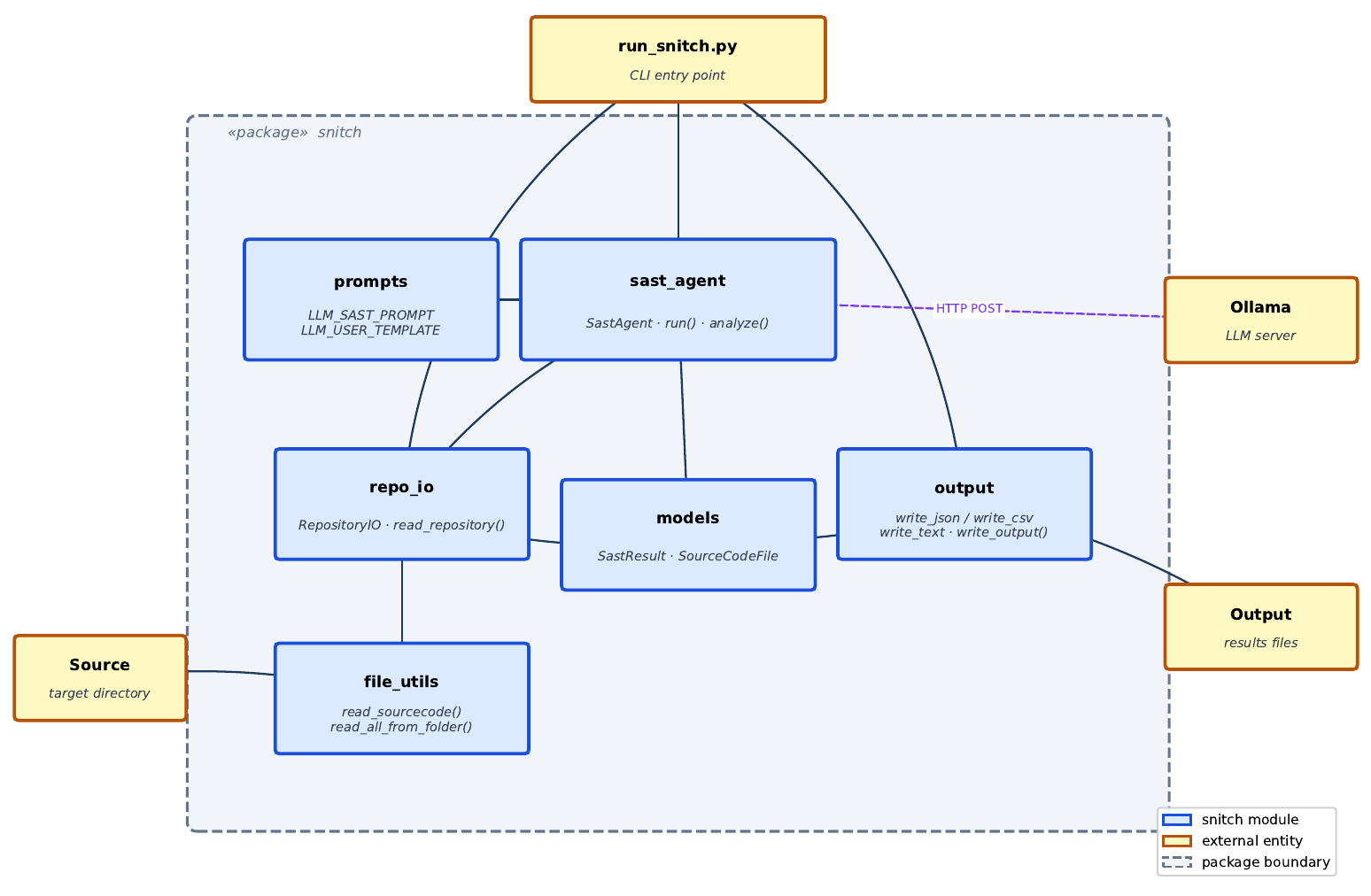}
    \caption{Snitch Component Diagram}
    \label{fig:snitchComponentDiagram}
\end{figure}

The next step in our experiment was the creation of a local LLM agent, which we named \textbf{Snitch} \cite{yohn_dyohnsnitch_2026}. Snitch is a Python-based agent application which leverages Meta's Ollama large language model runner to host LLMs on a user's local machine. A component view of Snitch is shown in Figure~\ref{fig:snitchComponentDiagram}. Ollama offers many models to choose, from general-purpose agents to coding-specific and beyond. For our experiment, we selected three models to power our agent: Google's \textit{gemma3}, Meta's \textit{llama3.1}, and Alibaba's \textit{qwen2.5}. Gemma is advertised as a coding specialist. Llama is presented as a general-purpose LLM. Qwen is intended for reasoning, coding, and debugging, but may be used more generally. Snitch consists of several Python modules which work together to recursively examine the contents of a user-specified directory. To keep this experiment manageable, only Python source code files are considered for scanning purposes; all others are skipped. 

Regardless of the underlying model used, Snitch relies on a fixed LLM prompt:

\vspace{1em}
\begin{quote}
\begin{verbatim}
You are an expert Cybersecurity engineer. Your specialty is in the realm
of Static Application Security Testing (SAST). You excel at examining source
code and identifying SAST problems such as security vulnerabilities,
injection flaws (SQL, command, etc.), cross-site scripting, hardcoded 
secrets/credentials, broken or misconfigured access controls, memory 
management issues (buffer overflows, memory leaks, etc.), cryptographic 
weaknesses, insecure framework/API misuse, and insecure path traversals.

Given a filename and its code contents, you MUST identify any SAST problems. 
If found, generate a summary of:
1. Where the problem was found (in `filename:codeStartLine-codeEndLine` 
    format)
2. What the problem is
3. Why it is a problem
4. What actions can be taken to eliminate or mitigate the problem

Output generated for each file MUST be written to a JSON object with fields:
- "where": string   (the output of #1)
- "what": string    (the output of #2)
- "why": string     (the output of #3)
- "fix": string     (the output of #4)

If you are unable to process a file, the JSON for the file should place the
name of the file as the value for the "where" key, and use "skipped" for 
the remaining key values.
\end{verbatim}
\end{quote}
\vspace{1em}

To begin the experiment, Bandit was used to scan the three selected open source tools. The resulting output formed the baseline of future comparisons and served as the reference against which all subsequent Snitch results were evaluated. Next, an Ollama model was selected for Snitch, and Snitch was then used to examine each of the three selected open source tools, with the output captured as a JSON file. The Snitch scans were repeated on each source code repository target, with each scan using one of the remaining models. In total, three baseline (Bandit) scans were collected, as well as nine agentic (Snitch) scans (three for each model).

In our evaluations of Snitch's output and comparison to Bandit, we primarily looked for issues that Snitch may have caught that Bandit missed, or vice-versa. We also considered false-positives from both scanning tools; note that Snitch's false-positives can be construed as hallucinated security threats, in addition to simply being issues that the prompts implied but did not explicitly state should be looked for. We made no attempt to quantify the confidence with which Snitch made a prediction, or the severity of any of its predictions, nor did we attempt any cross-referencing with known CVE definitions. We considered such efforts to be out of scope, in the sense that what we are explicitly evaluating is purely whether a generalized agent is capable of reliably and statically identifying well-known common vulnerabilities that would inevitably be present in publicly-available code used to train the target models. Items which Bandit has identified as medium or high severity are considered to be the most critical issues which Snitch aspires to flag.

Several metrics were calculated to aid in the empirical comparison of the various models to the baseline tool. First, a recall metric for each model was calculated, such that:
\begin{align*}
    &M = \text{Medium severity} \\
    &H = \text{High severity} \\
    &findings = (\text{number of M or H Bandit detections also found by a model}) \\
    &total = (\text{total Bandit M or H findings for the given repository}) \\
    &recall = findings \ / \ total
\end{align*}
Note that $findings$ counts both full and partial matches, with full matches being worth one point, partial matches worth 0.5 points, and no match as zero points. A Bandit issue is considered to be "missed" (not matched) if all three models fail to detect it.

A second important metric to calculate is the false-positive (FP) rate for each model. Snitch is considered to have produced a false-positive when it identifies a potential issue which both was not flagged by Bandit and which is clearly, per human judgment, a contextual misunderstanding (e.g., flagging the use of environment variables in a development configuration as a security violation). The false-positive rate is calculated as:
\begin{align*}
    FP_{rate} = (\text{manually-flagged FP count}) \ / \ (\text{number of potential FPs found}) \\
\end{align*}

Putting these two together, we can calculate a composite score for each model, where
\begin{align*}
    composite = recall \ - \ FP_{rate}
\end{align*}
Note that $FP_{rate}$ ranges from $-1$ to $1$, with higher (more positive) values being better. These calculations form the direct empirical basis of comparison to the accepted "source of truth", Bandit, whose scores are considered to be fully correct.

\begin{figure} [hbtp]
    \centering
    \includegraphics[width=0.8\linewidth]{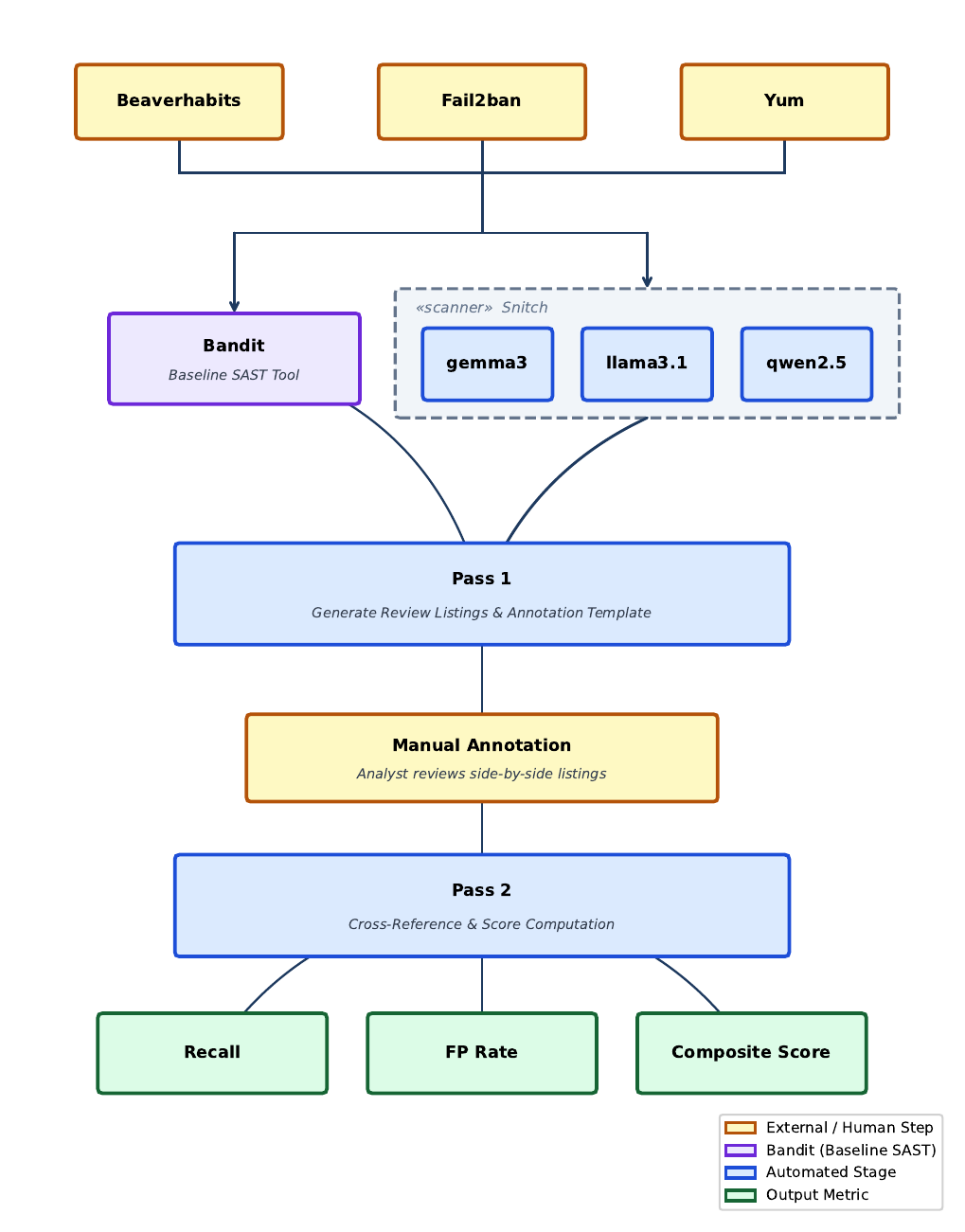}
    \caption{The Experimental Pipeline Employed}
    \label{fig:experimentalPipeline}
\end{figure}

A graphical flowchart of the experimental methodology described in this section can be viewed in Figure~\ref{fig:experimentalPipeline}.



\section{Results} \label{sec:results}

We begin presentation of empirical results with a simple tally of the number of files and lines of code scanned per repository, as seen in Table~\ref{tab:tallyFilesAndLOCs}. Bandit and each variant of Snitch was applied to the number of items listed.

\renewcommand{\arraystretch}{1.5}
\begin{table}[htbp]
    \centering
    \caption{Statistics of Files and LOCs Scanned on Each Run}
    \begin{tabular}{lcr}
         Repository & Python Files & Total Lines of Code (LOC) \\
         \toprule
         Beaverhabits & 71 & 9,124 \\
         Fail2ban & 81 & 34,569 \\
         Yum & 72 & 58,123 \\
         \midrule
         TOTAL & 224 & 101,816 \\
         \bottomrule
    \end{tabular}
    \vspace{0.5em}
    \label{tab:tallyFilesAndLOCs}
\end{table}
\renewcommand{\arraystretch}{1}

Next, we present the empirical results of the Bandit baseline. Running Bandit against the three target repositories identified numerous potential vulnerabilities with low, medium, or high severity. The detection counts for each severity level can be seen in Table \ref{tab:bandit_severity}. As discussed in Section \ref{sec:methodology}, detections that Bandit classified as low severity were discarded for the purposes of this experimental comparison.

\renewcommand{\arraystretch}{1.5}
\begin{table}[htbp]
    \centering
    \caption{Bandit findings by severity per repository}
    \begin{tabular}{lccccr}
        \toprule
        Repository & LOW & MEDIUM & HIGH & TOTAL & TOTAL - LOW\\
        \midrule
        Beaverhabits & 142 & 2 & 2 & 146 & 4 \\
        Fail2ban & 21 & 32 & 5 & 58 & 37\\
        Yum & 6 & 0 & 0 & 6 & 0\\
        \bottomrule
    \end{tabular}
    \vspace{0.5em}
    \label{tab:bandit_severity}
\end{table}
\renewcommand{\arraystretch}{1}

\begin{figure}[htbp]
    \centering
    \includegraphics[width=0.8\linewidth]{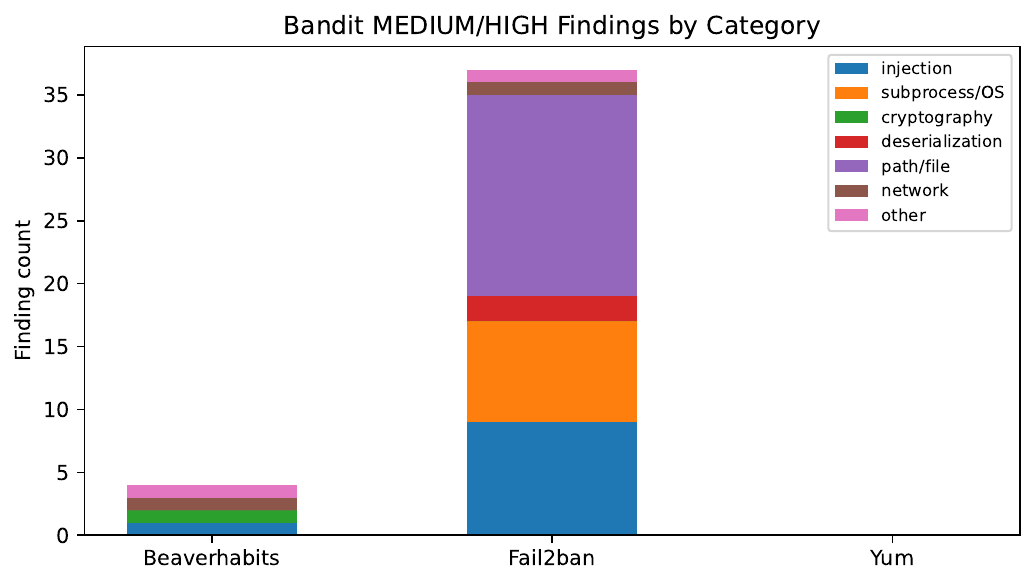}
    \caption{Bandit Categorical Findings Per Target Repository}
    \label{fig:bandit_mh_category_distribution}
\end{figure}

The types of issues identified by Bandit can be further classified into vulnerability categories. As expected, the Yum repository is mature and secure. After exclusion of low severity issues, Bandit detected no other potential issues in Yum. The raw counts of categorical vulnerabilities per repository is presented in Table \ref{tab:bandit_mh_categories}. The relative distribution of medium/high severity vulnerabilities per repository is most clearly visible in Figure \ref{fig:bandit_mh_category_distribution}. Beaverhabits exhibited a rather even distribution across four categories: injection, network, cryptography, and other. Fail2ban spanned six categories, with most issues falling into one of three categories: path/file, subprocess/OS, and injection. In general, many of the high-level issues were overlapping, with the most common being regularly starting processes and subprocesses with a shell. A common thread among all of these repositories was the improper parsing of XML data. Weak hashes are also common. Beaverhabits has more unique issues, most notably risky tarball extraction. Medium severity issues across the three repositories came down to regular use of \texttt{exec}, function calls with shell, open audit URL, and use of the marshal module with deserialization. Less frequently, insecure usage of temp file/directory was detected. A full accounting of the raw number of detections for all agent variants and Bandit can be seen in Table~\ref{tab:totalIssuesDetectedPerTool}.

\renewcommand{\arraystretch}{1.5}
\begin{table}[htbp]
    \centering
    \caption{Bandit MEDIUM/HIGH findings grouped by category}
        \begin{tabular}{lccr}
            \toprule
            Category & Beaverhabits & Fail2ban & Yum \\
            \midrule
            injection & 1 & 9 & 0 \\
            subprocess/OS & 0 & 8 & 0 \\
            cryptography & 1 & 0 & 0 \\
            deserialization & 0 & 2 & 0 \\
            path/file & 0 & 16 & 0 \\
            network & 1 & 1 & 0 \\
            other & 1 & 1 & 0 \\
            \bottomrule
        \end{tabular}
    \vspace{0.5em}
    \label{tab:bandit_mh_categories}
\end{table}
\renewcommand{\arraystretch}{1}

\renewcommand{\arraystretch}{1.5}
\begin{table}[htbp]
    \centering
    \caption{Issues Detected Per Agent/Tool}
    \begin{tabular}{lllll}
         Repository & llama3.1 & gemma3 & qwen2.5 & Bandit  \\
         \toprule
         Beaverhabits & 56 & 56 & 54 & 146 \\
         Fail2ban & 25 & 25 & 22 & 58 \\
         Yum & 17 & 18 & 20 & 6 \\
         \bottomrule
    \end{tabular}
    \vspace{0.5em}
    \label{tab:totalIssuesDetectedPerTool}
\end{table}
\renewcommand{\arraystretch}{1}

The Snitch agent was run against all three target repositories three times, with each run utilizing a different underlying LLM. Bandit was used as the baseline reference tool rather than an absolute ground truth, with the general expectation that at least one of the three models used would be able to substantively reproduce the fundamental insights provided by Bandit. That was not to be the case. While many potential issues were identified, Snitch consistently underperformed compared to Bandit across the evaluated metrics. 

Snitch output files were noisy and sometimes unfocused as compared to Bandit. A notable challenge was that Snitch was more likely than not to hallucinate the \texttt{where} component of each detection. The purpose of the \texttt{where} component was to identify the file name and line number range associated with each detected issue. In practice, this information was frequently inaccurate and often an outright hallucination, referring to non-existent files or code locations. Fortunately, the name and path of each scanned file was also captured for each detection with greater accuracy, allowing us to move forward with the analysis.

To manage the unstructured output generated by Snitch, a hybrid analysis process was developed where a secondary LLM, Claude, was used to generate a processing script to cross-reference the nine generated Snitch output files versus the three Bandit source of truth output files. Potential matches were flagged for human review, and all non-matching detections (those generated by Snitch but not by Bandit) were routed to secondary files for additional triage and human inspection. During human review, partial credit was given for Snitch correctly identifying a file as having some type of problem (as determined by Bandit), making this a permissive and generous evaluation. All potential matches were manually cross-referenced with the original target source code to validate and verify correct detections, contextual misunderstandings, or outright hallucinations. It is important to note that there is significant overlap with contextual misunderstandings and pure hallucinations. That is, Snitch was found to suffer from a limited context window, which often caused it to misinterpret common software architectural and development patterns as security vulnerabilities. For example, a common practice in modern Python development is the use of a so-called \texttt{.env} file for local development secrets, which are then consumed by Python's \texttt{dotenv} or \texttt{pydantic} libraries; Snitch consistently identified this pattern as hard-coded secrets leaking to production, despite the obvious intent to use environment variables or \texttt{Pydantic} class objects rather than on-disk secrets.

Across all of the LLMs leveraged, it was very common for the \texttt{dotenv} pattern to be flagged as an error (inclusion of hard-coded secrets). Each model suffered from a non-trivial number of false positive (FP) results, detections that flagged non-issues (benign code) as security vulnerabilities. These FPs took various forms, such as licenses and comments as errors, purported SQL injections when processing Python's string "dunder" conversion method, Emacs and Vi command headers flagged as non-functional/useful, misidentifying standard library constants as hard-coded secrets (e.g., \texttt{calendar.MONDAY}), and more. The gemma3 model exhibited the highest number of FPs, followed by llama3.1, with qwen2.5 having the fewest.

Coupled with the propensity for false positives was a disastrously low accuracy. A substantial proportion of the detections generated by the evaluated models were determined to be FPs. There were a few bright spots, though they are unverifiable since our source of truth did not register the issues. In Beaverhabits, llama3.1 flagged an insecure use of Python's \texttt{requests.get()} with unvalidated user input, suggesting the \texttt{urllib} library for better parsing and validation, while qwen2.5 identified two new issues: an insecure path traversal and potential injection flaw, both in core modules. In Fail2ban, llama3.1 identified two potential issues: a SQL injection and a vulnerability to Cross-Site Scripting (XSS) attack. Also in Fail2ban, qwen2.5 discovered one implementation problem (inheriting from a base class without implementing all functionality), and two potential vulnerabilities: insecure/incomplete exception handling and insecure command execution.



\begin{figure}
    \centering
    \includegraphics[width=\linewidth]{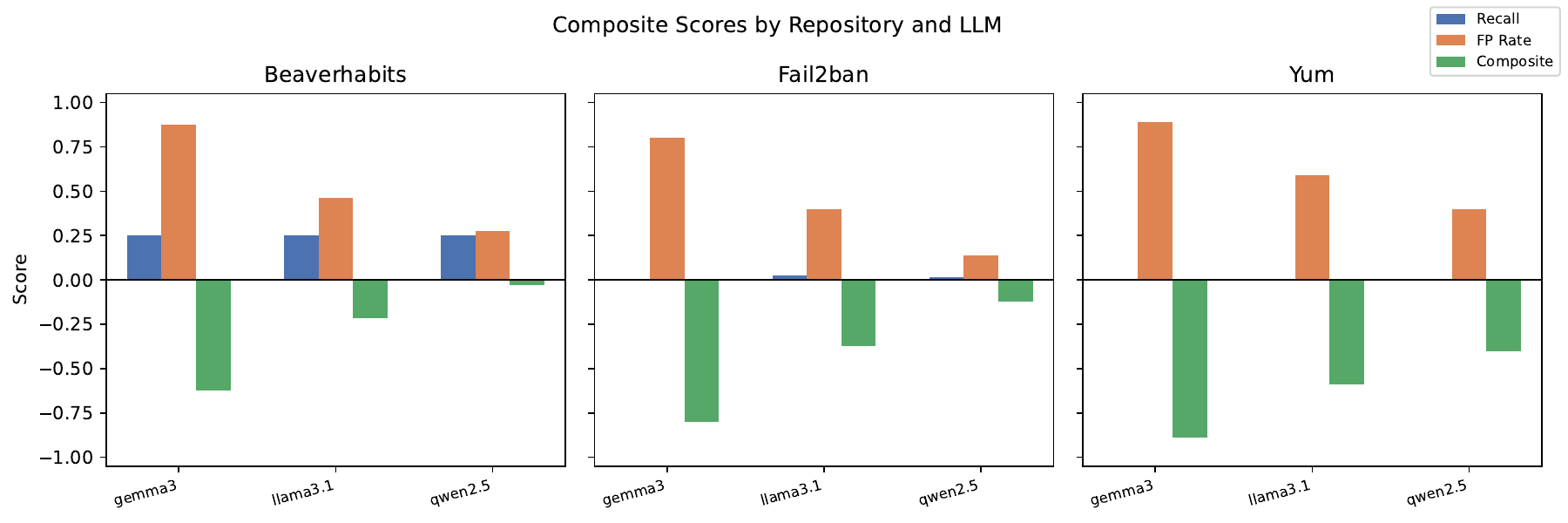}
    \caption{Recall, False Positives (FP), and Composite scores per model}
    \label{fig:compositeScores}
\end{figure}

Composite scores can be graphically compared by viewing Figure \ref{fig:compositeScores}. For a more complete comparative analysis of recall, FP, and composite scores, see Table \ref{tab:composite_scores}. Among the evaluated models, \textbf{qwen2.5} achieved the strongest overall performance. However, this represents relative superiority among under-performing models rather than a strong indicator of practical suitability.

\begin{table}[htbp]
    \centering
    \caption{Composite scores per repository and LLM (recall $-$ FP rate; best per repo in bold)}
    \begin{tabular}{l rrr rrr rrr}
        \toprule
        & \multicolumn{3}{c}{gemma3} & \multicolumn{3}{c}{llama3.1} & \multicolumn{3}{c}{qwen2.5} \\
        \cmidrule(lr){2-4}\cmidrule(lr){5-7}\cmidrule(lr){8-10}
        Repository & Recall & FP Rate & Composite & Recall & FP Rate & Composite & Recall & FP Rate & Composite \\
        \midrule
        Beaverhabits & 0.25 & 0.88 & -0.62 & 0.25 & 0.46 & -0.21 & 0.25 & 0.28 & \textbf{-0.03} \\
        Fail2ban & 0.00 & 0.80 & -0.80 & 0.03 & 0.40 & -0.37 & 0.01 & 0.14 & \textbf{-0.12} \\
        Yum & 0.00 & 0.89 & -0.89 & 0.00 & 0.59 & -0.59 & 0.00 & 0.40 & \textbf{-0.40} \\
        \bottomrule
    \end{tabular}
    \vspace{0.5em}
    \label{tab:composite_scores}
\end{table}

There remains one final point of comparison worthy of presentation: execution speed. We did not set out to consider, measure, or compare the execution speeds of the various Ollama models leveraged by our agent. For that reason, no experimental planning or metric gathering was done along that avenue of inquiry. However, we could not help but notice that the execution of a local agent for the purposes of SAST scanning was dramatically prohibitive. Where a rule/pattern-based system such as Bandit is capable of executing these scans in well under one minute per repository, Snitch scans routinely took one to four hours to complete, leading us to execute several runs overnight. For the purposes of this experiment, that is nothing more than an inconvenience. In the real world, however, such time spans could cause greater overhead. That said, a rough estimate of agent time per scan is presented in Table~\ref{tab:estimatedTimePerRun}. Each LLM call through Ollama on a local server takes approximately 30-90 seconds per file depending on file size. In short, there were 224 files to be scanned by each of three models, yielding a total of 672 LLM calls. At approximately 60 seconds per file per LLM call, we experienced a total runtime in the neighborhood of 10-12 hours (at least 672 minutes) across all three agent/model combinations. Given the lack of additional empirical evidence for this metric, further speculation will be provided in Section~\ref{sec:discussion}.

\renewcommand{\arraystretch}{1.5}
\begin{table}[htbp]
    \centering
    \caption{Time Per Repo Scan for Each Agent}
    \begin{tabular}{lcr}
         Repository & Lines of Code (LOC) & Estimated Time Per Run Per Model \\
         \toprule
         Beaverhabits & 9,124 & ~1 hour \\
         Fail2ban & 34,569 & ~2-3 hours \\
         Yum & 58,123 & ~3-4 hours \\
         \bottomrule
    \end{tabular}
    \vspace{0.5em}
    \label{tab:estimatedTimePerRun}
\end{table}
\renewcommand{\arraystretch}{1}

Creation of Snitch was undertaken by multiple researchers, and all researchers collaboratively selected the baseline comparison tool, Bandit, as well as the target Python repositories to be scanned for evaluation. Baseline and agent test results were generated by only one researcher. The application of hybrid analysis via Claude resulted in the generation of cross-referenced files of flagged discrepancies for further human review. However, multiple researchers manually reviewed the original result files and flagged cross-reference files for matches, false positives, and other anomalies. All researchers participated in further discussion and review of independent findings from study of the outputted result files. We believe that our cross-collaboration during experimental design and review has ensured our findings were consistent with the evidence gathered.

Unlike previous methods, we have attempted to use off-the-shelf general-purpose LLMs to build an agent capable of fulfilling a specialized task. We provided no tailoring or intricacies to hide behind custom implementations or special cases. We believe that the empirical results speak for themselves: The evaluated models exhibited very low recall and high false-positive rates throughout the experiment. It is certainly possible that closed source or specialist models, coupled with improved prompting strategy, may provide substantially better results. However, we argue that our results are significant even in the context of more sophisticated experiments, such as the those presented in Section~\ref{sec:relatedWork}. The magnitude of the observed experimental results suggests that similar outcomes may be observed in comparable experimental settings. Additional studies involving alternative models, prompts, and repositories are needed to assess the generalizability of these findings, as agentic LLM-powered systems may yet prove to be a sea change in how future systems are built to solve real world problems. Within the scope of this study, hallucinations and explainability limitations remained significant challenges.

\section{Discussion} \label{sec:discussion}

The results indicate that Bandit outperformed the evaluated AI-based systems in both effectiveness and execution speed. The results Bandit gave were consistent, clear, and provided results that give agency to the users, who may have a better understanding of the workings of the system than the pattern-matching methods of Bandit. Bandit also has the privilege of being incorrect and responding with lower confidence, allowing users to question the need to remove or change flagged vulnerabilities. By ranking these results Bandit also allows users to attack from the top-down, and fix more critical infrastructure sooner. Bandit is a simple system, but straightforward to implement and easy to understand, with a high degree of explainability and traceability.

The evaluated LLMs frequently presented their findings with high linguistic confidence, even when the underlying detections were inaccurate. The LLM responses, in this experimental context, were regularly incorrect, raising questions of validity and creating potential for even greater security threats by changing things that may not require changing. In addition, without a massive summary of the results or quantitative summaries of one result, the undertaking of security analysis becomes tedious, possibly wasting potential resources on minor or relatively unimportant issue. Even if the agentic AI was prompted to give the most important measurements, the hallucinations regarding location make the program nearly worthless, making the knowledge that there may be a critical issue without knowing where it is unhelpful. LLMs already suffer from a lack of explainability, and it is difficult to see the utility in giving end-users more reasons to distrust the output of such systems. In short, we replicate Khan et al.'s \cite{khan_ai-driven_2025} findings of the unexplainability problem and Wang et al.'s \cite{wang_enhanced_2026} findings that off-the-shelf models do not perform well on specialized tasks, and that hallucinations are omnipresent and contribute to developer distrust of LLM-powered systems.

Setting aside initial impressions, it is clear that the rule/pattern-based approach of a structured tool like Bandit has significant advantages versus the open-ended generative stream of a large language model. The output from a structured tool is succinct and lends itself well to programmatic parsing. Despite our best attempts, we could not achieve similar structure from our agent, regardless of the underlying LLM. The agent certainly attempted to conform to its prompt guidance, and did, in fact, output JSON objects. However, the rampant hallucinations for the \texttt{where} attribute made it simply impossible to rely upon fully non-human processing, replicating a key finding of Xu et al. \cite{xu_seek_2026}. Ironically, this worked to our advantage, as it forced per-entry investigation and validation of the Snitch results, though this necessitated use of a non-deterministic processing/sorting algorithm to divide Snitch's output into manageable files. Our manual examinations give us confidence that the limitations we have discovered are not simply an artifact of a misguided prompt, but rather an inescapable conclusion of a generative system, incentivized to guess and deprived of the external information (CVE definitions) necessary to build a pattern-based system. One wonders what greater good would be served by gathering all the requirements of Bandit and then hoping we miraculously generated a competitive program if we simply toyed with the prompt long enough to empirically prove full compliance and correctness. Our work hints that such a desire appears to be forever hampered by the inherent hallucinations of LLMs, as first discovered by multiple related works \cite{wang_enhanced_2026, xu_seek_2026, klemmer_using_2024, nethala_cyber_2025}.

The results suggest that the evaluated general-purpose open-source LLMs are not well suited to static application security testing under the conditions examined in this study. The quantitative results consistently indicate limited effectiveness across the evaluated models. False-positive rates ranging from 49\% to 90\% indicate substantial limitations in detection reliability. Recall values did not exceed 0.25 in any experimental condition, indicating considerable room for improvement, whether with real data (if any remains available on the internet) or synthetic data (mimics of copies of guesses). These "general-purpose" open source tools are not truly general-purpose, and without such models agent-based systems are limited in their applicability. Although agentic AI is usable for generating code, predicting patterns, and automating more tedious tasks, it still has tremendous room for growth, particularly in the realm of specialized tasking such as cybersecurity. While this study provides empirical evidence regarding agentic SAST scanning, additional research is needed to fully address the gap identified by Al-Shammare et al. \cite{al-shammare_evaluating_2025}.

We must also consider the possible conditions which made \textbf{qwen2.5} the most performant model-engine that we examined. All of the models selected for testing were purported to be general-purpose. However, qwen2.5  produced fewer findings on Beaverhabits and Fail2ban but more on Yum, suggesting higher precision rather than simply lower recall. We believe there are three possible explanations for that outcome:
\begin{enumerate}
    \item \textbf{Reasoning Orientation:} qwen2.5 was trained by Alibaba with strong emphasis on structured reasoning and code understanding. It applies more deliberate judgment before flagging, reducing false positives on simpler repositories.
    \item\textbf{Training Data Composition:} qwen2.5 is trained on a large mutinational corpus with significant open source code coverage, potentially giving it better calibration for distinguishing real vulnerabilities from style issues.
    \item \textbf{Context Window Utilization:} Yum is the largest repository at 58,123 lines. qwen2.5 found more findings there (20 vs 17 vs 18), suggesting it handles larger file contexts more effectively than the other models. 
\end{enumerate}

While we endeavored to create a focused and thorough "persona" prompt, the same prompt was used for each agent run. This was done to provide comparative consistency and implicitly test whether "general-purpose" LLMs are, indeed, general-purpose. Our testing findings indicate that the evaluated LLMs performed poorly on a specialized task when operating under a fixed prompting strategy. It is certainly possible that prompt engineering improvements might meaningfully change the results, but we leave such attempts as future work. Indeed, we believe such prompt engineering approaches should be investigated, especially in light of the clear failure outcomes we have empirically proven herein.

Finally, let us consider the computational and time complexity of generative systems. LLM models take longer amounts of time to do tasks than traditional programs like Bandit, and even if time/space complexity was equivalent then there would still be the inherent non-determinism of such systems to consider. When it is something like cybersecurity, the results need to be more consistent than what an agentic AI is currently capable of achieving. For token-based intelligent systems such as LLMs, increased execution time is not simply a delay or annoyance, it might very well be a catastrophic unexpected cost if using a pay-per-usage LLM. Such impacts are both undesirable and untenable. Consequently organizations are generally unlikely to accept substantial safety and financial risks solely for the purpose of adopting or promoting the newest software or intelligent systems.

At best, there are practical uses for Agentic and common AI in cybersecurity, such as hybrid systems. For example, we found several instances where the Snitch agent discovered potential issues that Bandit missed. Perhaps a wiser use of these intelligent systems would be as supplements to known and tested classical scanning tools. There would, of course, be the issue of accuracy and hallucination. However, a narrowed focus, where only newly discovered issues were investigated, might be fruitful and increase overall system security posture; this is perhaps an interesting research direction that could build off the work we have presented herein. Currently, though, traditional rule-based tools appear to be the best when analyzing code for cybersecurity vulnerabilities, at least as compared to open source, locally hosted LLM agents. 

\section{Threats to Validity} \label{sec:threatsToValidity}

\subsection{Construct Validity}
Our operationalization of agent capability is subject to two related limitations. First, we restricted our investigation to open-source, general-purpose large language models paired with a fixed prompt. Closed-source models are likely to be significantly more powerful, though they may carry high usage costs; with proper funding, a more thorough investigation could be undertaken. Second, and relatedly, the prompt itself was not modified or optimized, which means empirical results may not reflect the full potential of the approach.

\subsection{External Validity}
The generalizability of our findings is limited in two ways. Our choice of scannable target repositories may not be representative of the types of codebases an agentic system would commonly examine; a larger variety of targets may provide a more complete picture of possible outcomes. Additionally, our reliance on Bandit as a single source of ground truth may be a weakness. The choice of Bandit was made for technical and financial reasons, and given more time and resources it would be prudent to expand our basis of comparison to include other well-known and validated tools.

\subsection{Internal Validity}
Our experimental design was lacking in several key areas, including scan timing metrics, assessment confidence, issue severity level, and cross-referencing with known vulnerability databases. It may be the case that most detections made by Snitch were of low confidence or severity and should therefore have been discarded during evaluation. There is also some variant of the single-annotator threat to consider. A separate LLM was used to generate cross-referenced files of flagged issues, which were then manually examined by one researcher and cross-validated by the remaining researchers. While we believe we have effectively mitigated any potential issues arising from this hybrid analysis approach, one can never be too careful. Our data and agent implementation are freely available \cite{yohn_dyohnsnitch_2026} to assist other researchers in refuting, validating, or expanding upon our work.

\subsection{Conclusion Validity}
Finally, our agent design, including the prompt, may have been far too na\"{i}ve to support strong conclusions. Further iteration, experimentation, and testing of the software design may have yielded significantly improved results, and observed performance may therefore reflect implementation weakness rather than the ceiling of the approach.

\section{Conclusion} \label{sec:conclusion}
In this study, we created a Python-based AI Agent and tested its efficacy when powered by three different Ollama-hosted general-purpose open source models: gemma3, llama3.1, and qwen2.5. We assess each agent's performance using precision, recall, false positives, and a calculated composite score combining all metrics, against the baseline performance of an existing, vetted SAST tool, Bandit. We find that our agent’s performance was substantially lower than that of the baseline tool, regardless of the model engine selected. The evaluated general-purpose models produced a high number of false positives, while simultaneously failing to detect standard issues that Bandit unveiled easily. Hallucinations and context-related errors were frequently observed, and the evaluated open-source models were not effective for the SAST task under the experimental conditions used in this study. Additionally, agent processing times are dramatically longer than those exhibited by Bandit, yielding an alternative solution that is untenable from both a reaction time and potential costs perspective.


\bibliographystyle{IEEEtran}





\bibliography{AgenticCyberSecScanning}


\end{document}